\documentclass[12pt]{iopart}

\usepackage{harvard}
    \citationmode{abbr}
\usepackage{graphicx}  
\usepackage{subfig}
\usepackage{booktabs}
\usepackage{xcolor}
\usepackage{comment}
\usepackage{siunitx}

\begin{document}

\title[GEANT4 simulations of HTM RV]{GEANT4 simulation of a new range verification method using delayed $\gamma$ spectroscopy of a $^{92}$Mo marker}

\author{E Kasanda$^1$, C Burbadge$^1$, V Bildstein$^1$, J Turko$^1$, A Spyrou$^{2,3}$, C H\"{o}hr$^{4,5}$ and D M\"{u}cher$^{1,4}$}

\address{$^1$ Department of Physics, University of Guelph, 50 Stone Rd E, Guelph, Ontario, Canada N1G 2W1}
\address{$^2$ National Superconducting Cyclotron Laboratory, Michigan State University, East Lansing, Michigan 48824, USA}
\address{$^3$ Department of Physics and Astronomy, Michigan State University, East Lansing, Michigan 48824, USA}
\address{$^4$ TRIUMF, 4004 Wesbrook Mall, Vancouver, British Columbia, Canada V6T 2A3}
\address{$^5$ Department of Physics and Astronomy, University of Victoria, 3800 Finnerty Road, Victoria, British Columbia, Canada, V8P 5C2}

\ead{dmuecher@uoguelph.ca}
\vspace{10pt}
\begin{indented}
\item[]November 2019
\end{indented}

\begin{abstract}
In this work, we propose a novel technique for in-vivo proton therapy range verification. This technique makes use of a small hadron tumour marker, $^{92}$Mo, implanted at a short known distance from the clinical treatment volume. Signals emitted from the marker during treatment can provide a direct measurement of the proton beam energy at the marker's position. Fusion-evaporation reactions between the proton beam and marker nucleus result in the emission of delayed characteristic $\gamma$ rays, which are detected off-beam for an improved signal-to-noise ratio.
In order to determine the viability of this technique and to establish an experimental setup for future work, the Monte Carlo package GEANT4 was used in combination with ROOT to simulate a treatment scenario with the new method outlined in this work. These simulations show that analyzing the intensity of delayed $\gamma$ rays produced from competing reactions yields a precise measurement of the range of the proton beam relative to the marker, with sub-millimetre uncertainty.\\\\

\noindent{\it Keywords\/}: proton therapy, range verification, GEANT4, simulation, tumour marker, $\gamma$ spectroscopy

\end{abstract}
\section{Introduction}
Despite the steady decline in cancer-related death rates for the past 30 years, cancer remains the leading cause of death in the developed world, affecting nearly 1 in 2 people in their lifetime \cite{cancersociety}. Approximately 50\% of all cancer patients undergo radiation therapy as a part of their treatment \cite{Barton2014,Tyldesley2011}, often in combination with other methods such as surgery and chemotherapy. In the past few decades, proton therapy (PT) has gained popularity as a method of radiation therapy because of the advantage it offers in terms of dose distribution compared to conventional radiation therapy \cite{Knopf2013,mcgowan2013,Wilson1946}.

Conventional radiation treatment methods make use of $\gamma$ rays to deposit an ionizing radiation dose to a targeted region of tissue. Exponential attenuation of the photons delivered through this type of treatment results in a significant dose being delivered to healthy tissue along the beam path, both before and after the targeted region. Modern techniques using intensity modulation and multiple beam angles allow treatment plans to compensate for this attenuation and minimize dose to healthy tissue. For heavy charged particles like protons, the dose deposition in matter is fundamentally different from that of photons \cite{mcgowan2013,Wilson1946}. The stopping power (energy absorbed per unit length) of tissue is described by the Bethe equation \cite{Bethe1930}, which states that the stopping power for a proton is approximately inversely proportional to the proton's energy \cite{Wilson1946}. Thus, protons deposit very little entry dose in tissue. Instead, the stopping power increases non-linearly with depth, reaching its maximum value near the end of the proton track. This results in an extremely sharp dose deposition profile with no exit dose \cite{mcgowan2013,Wilson1946}. The region of maximal dose deposition is referred to as the Bragg Peak, and its depth in tissue is dependent on the initial energy of the proton beam and the proton stopping power of the tissue \cite{Wilson1946}. The highly localized nature of this dose profile allows for precise irradiation of tumours with minimal dose delivered to the healthy tissue surrounding the targeted volume \cite{Knopf2013}. Furthermore, the high linear energy transfer (LET) of protons has been shown to result in an increased relative biological effectiveness (RBE) when compared with low-LET radiation such as photons \cite{Ilicic2018}. In combination, these factors make PT a promising method for the treatment of malignant tumours. It has been shown that PT has the potential to provide a highly localized and effective dose delivery with sub-millimetre precision \cite{Paganetti2012}.

Since the majority of the proton dose is deposited in the span of a few millimetres, it is crucial that the range of the beam inside the patient is precisely known in order to avoid accidental irradiation of healthy tissue with the Bragg Peak and, consequently, underdosing of the clinical target volume (CTV) \cite{mcgowan2013}. Range verification (RV) is an essential tool for characterizing the dose being delivered to the CTV and surrounding tissue during fraction delivery. As it stands, uncertainty in patient motion, tissue stopping powers and biological effectiveness contribute to an overall range uncertainty on the order of millimetres \cite{mcgowan2013}. To account for this, current PT planning target volumes (PTVs) include a safety margin of several millimetres, increasing the irradiated volume of healthy tissue in order to avoid underdosing the CTV \cite{Knopf2013}. 

While in principle, the lack of exit dose in PT could allow tumours to be targeted near organs at risk (OARs) \cite{mcgowan2013}, the range uncertainties are greater than in conventional therapy, which increases the risk of irradiating nearby OARs with the distal edge of the Bragg peak. To mitigate the risk of second malignancies, the axial alignment of the proton beam with the OAR is sometimes avoided, preferentially delivering dose to these tumours with the lateral edge of the beam as opposed to the much sharper distal edge \cite{verburg2014}. 

Without an exit dose to be measured for dose verification purposes, most current RV approaches in PT rely on the measurement of $\gamma$ rays produced from proton activation of tissue \cite{Parodi2007,parodi2018,Paganetti2015}. 

One of these approaches makes use of $\beta$-delayed positron annihilation events, which are responsible for the production of high intensities of 511 keV $\gamma$ rays in tissue along the beam path \cite{parodi2018,TOBIAS1977}. These $\gamma$ rays can be measured with PET detectors to produce a spatial reconstruction of the beam's path in the patient. However, because protons primarily deposit energy through electromagnetic interactions, the measured PET activation is not directly correlated with the dose distribution and must be compared to a Monte Carlo (MC) simulation in order to fully characterize the beam's range inside the patient \cite{parodi2018}. Since PET activation is not specific to any particular nuclear reaction, this MC simulation must take into consideration the precise tissue composition of the patient, obtained through a CT scan \cite{Parodi2005}. Uncertainties in the conversion of CT Hounsfield units (HU) to tissue composition and scarcity of experimental data for isotope production cross sections in a thick target both contribute to inaccuracies in the MC simulation \cite{Paganetti2015}.
When performed offline, PET RV measurements are taken several minutes after fraction delivery. It has been shown that this time delay can impact the range precision that can be achieved through this method due to the decay of a significant portion of the positron emitters before the PET scan is performed \cite{Paganetti2015}. To compensate for this, long acquisition times, typically on the order of 20-30 minutes, are generally required in order to obtain sufficient statistics for accurate range and dose calculation \cite{Parodi2007}. This long time delay leaves room for the blurring effects of biological decay and washout, which are difficult to predict due to the varying perfusion of different tissues \cite{Parodi2007}. Online PET is an alternative RV method in which the PET scanner is located inside the treatment room, incorporated with the proton beam ``nozzle'' and acquires data during fraction delivery in order to minimize the effects of the time delay present in offline measurements. High background $\gamma$ rates during fraction delivery and limited angle coverage due to spatial constraints are the main limiting factors of this method \cite{Paganetti2015,parodi2018}. The spatial resolution of a PET scanner is fundamentally limited by acollinearity and positron range, such that the theoretical limit of the spatial resolution for PET is a \SI{1.83}{\milli\meter} FWHM \cite{moses2011}. Typical clinical uncertainties in PET RV are on the order of 3-\SI{5}{\milli\meter} \cite{Paganetti2015}. 

Prompt $\gamma$ (PG) RV makes use of the multitude of prompt characteristic $\gamma$ rays which are emitted from nuclear interactions between the proton beam and tissue along the beam path. Due to the short time scale in which they are emitted, the acquisition time for this method is much shorter than that of PET, so the PG method has the potential to provide immediate feedback on treatment quality. The prompt aspect of the measurement allows PG RV to minimize the effects of biological washout and signal decay that must be considered in offline PET RV. In addition, the use of characteristic $\gamma$ rays as opposed to annihilation $\gamma$ rays minimizes the need to precisely estimate tissue composition \cite{verburg2014}. Several different imaging configurations have been investigated for PG RV. One basic setup makes use of passive collimation to ensure only $\gamma$ rays from a small region of the tissue are measured. Due to the high energy (up to 6 MeV) of the $\gamma$ rays being emitted at high intensities, it is crucial that the collimator is thick enough to sufficiently attenuate $\gamma$ rays originating from outside the region of interest \cite{verburg2014}. During treatment, the detector may either be focused on one spatial region, or scan the range of the beam in the patient. The resulting energy and timing distribution of the $\gamma$ rays produced during treatment is then compared to MC simulations in order to reconstruct the absolute range of the beam in a uniform water-equivalent phantom with an uncertainty of 1.0-\SI{1.4}{\milli\meter} \cite{verburg2014}. More recently, investigations of actively-collimated PG RV have been making use of a Compton camera, whose position-sensitive detectors are arranged such that the $\gamma$-ray angle of incidence can be reconstructed in 3D without the need of a collimator \cite{Hueso2016}. The abundance of $\gamma$ rays measured by a Compton camera can then be mapped as a function of their origin in the tissue in order to detect shifts in the range of the proton beam as small as \SI{2}{\milli\meter} \cite{Draeger2018}. This reconstruction is very complex and computationally intensive, and requires high statistics. In addition, the demanding energy, timing, and spatial resolution requirements of the Compton camera detectors make it a costly investment \cite{Draeger2018}. As with PET, all current PG RV methods rely on comparison to MC simulations for range and dose verification.


In this work, we propose a novel technique for in-vivo PT RV, hereafter referred to as HTM RV, that can be used in clinical practice. This technique involves the implantation of a small Hadron Tumour Marker (HTM) in the beam path, at a short known distance from the CTV. Signals emitted from the HTM during treatment can provide a direct measurement of the proton beam energy at the marker's position. The proximity of the HTM interaction site to the end of the beam's range allows for a higher-precision range calculation than could be achieved by measuring the beam energy outside the patient.

An HTM is a small metal marker composed of materials with a favourable response to proton activation. Factors considered in choosing a suitable HTM include the magnitude and energy dependence of the nuclear reaction cross sections, the half lives of the nuclei produced from these reactions, and the $\gamma$-ray energies emitted in the decay of these nuclei, as well as their respective intensities. Other relevant factors from a clinical perspective are the option to use natural materials as markers instead of isotopically enriched ones, as well as biological and long-term radiation toxicity of the marker. However, the primary focus of this work is to investigate the feasibility of this method with regard to accuracy and sensitivity. Biological aspects will be discussed in forthcoming publications. 

\begin{figure}[!ht]
  \centering
  \includegraphics[width=0.5\textwidth]{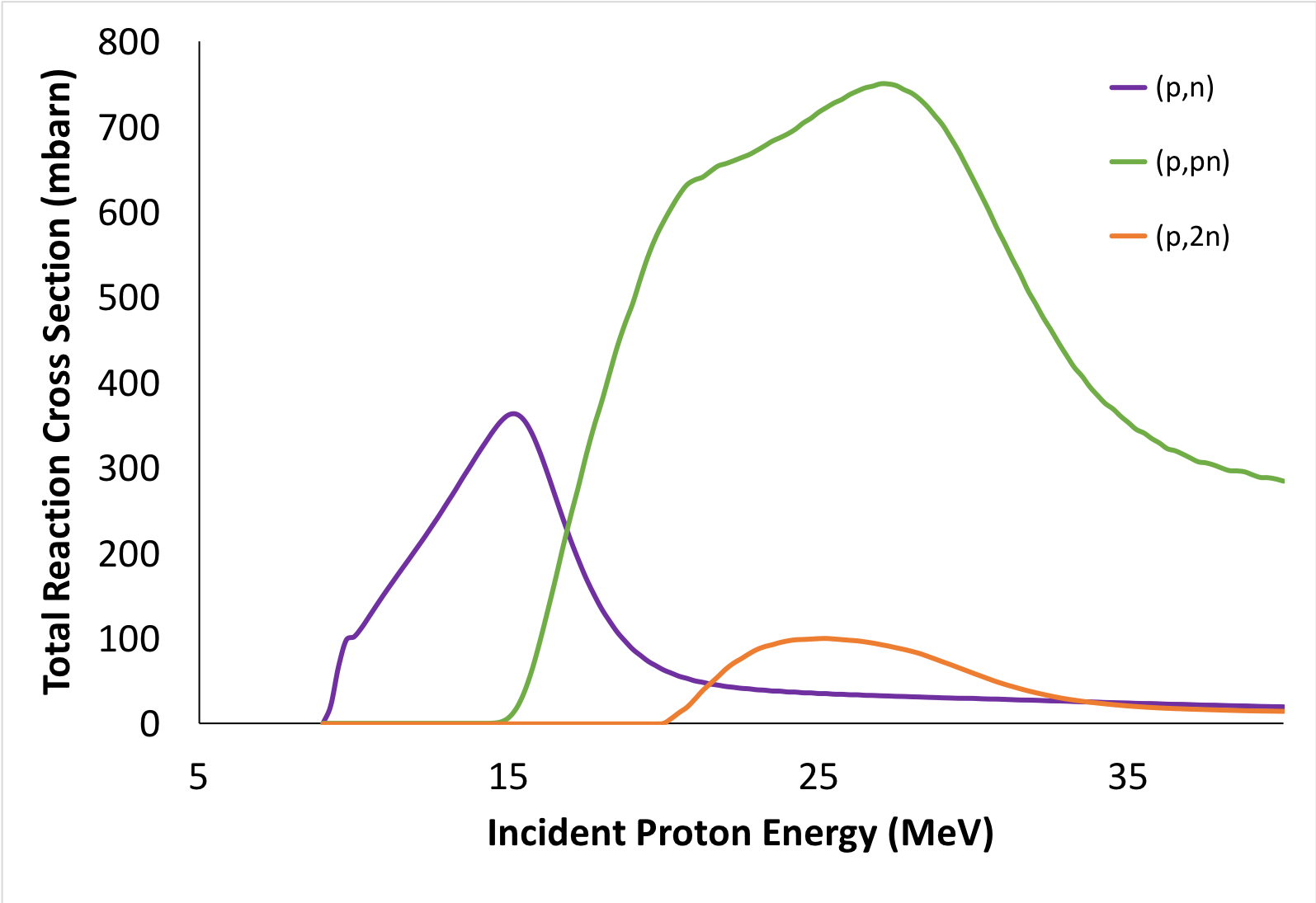}
  \caption{Plot of select fusion-evaporation reaction cross sections for $^{92}$Mo as a function of incident proton energy, obtained using TALYS software.}\label{fig:xsec-sketch}
\end{figure}
 
Fusion-evaporation reactions between the proton beam and the HTM during fraction delivery result in the production of stable and unstable nuclei. The marker isotope $^{92}$Mo was selected as a HTM candidate due to its response to proton activation. Figure \ref{fig:xsec-sketch} shows a plot of different reaction cross sections for $^{92}$Mo as a function of incident proton energy, obtained using the TALYS software (version 1.8) \cite{KONING2012}.  The cross sections for competing reactions tend to be maximized at different proton energies. This means that the intensity of the characteristic $\gamma$ rays of interest is correlated with the energy of the proton beam, and that the relative intensity of $\gamma$ rays produced from competing reactions can be used as a measurement of beam energy at the position of the marker, without the need to refer to MC simulations. An important advantage of measuring the proton energy at the marker location is that the heterogeneity of tissue stopping powers upstream of the marker no longer contributes to the range uncertainty, and the remaining range of the beam after the marker can be estimated with minimal uncertainty. The measurement of these $\gamma$ rays allows for sensitive in-vivo RV for PT fraction delivery.

\begin{figure}[!ht]
  \centering
  \includegraphics[width=0.9\textwidth]{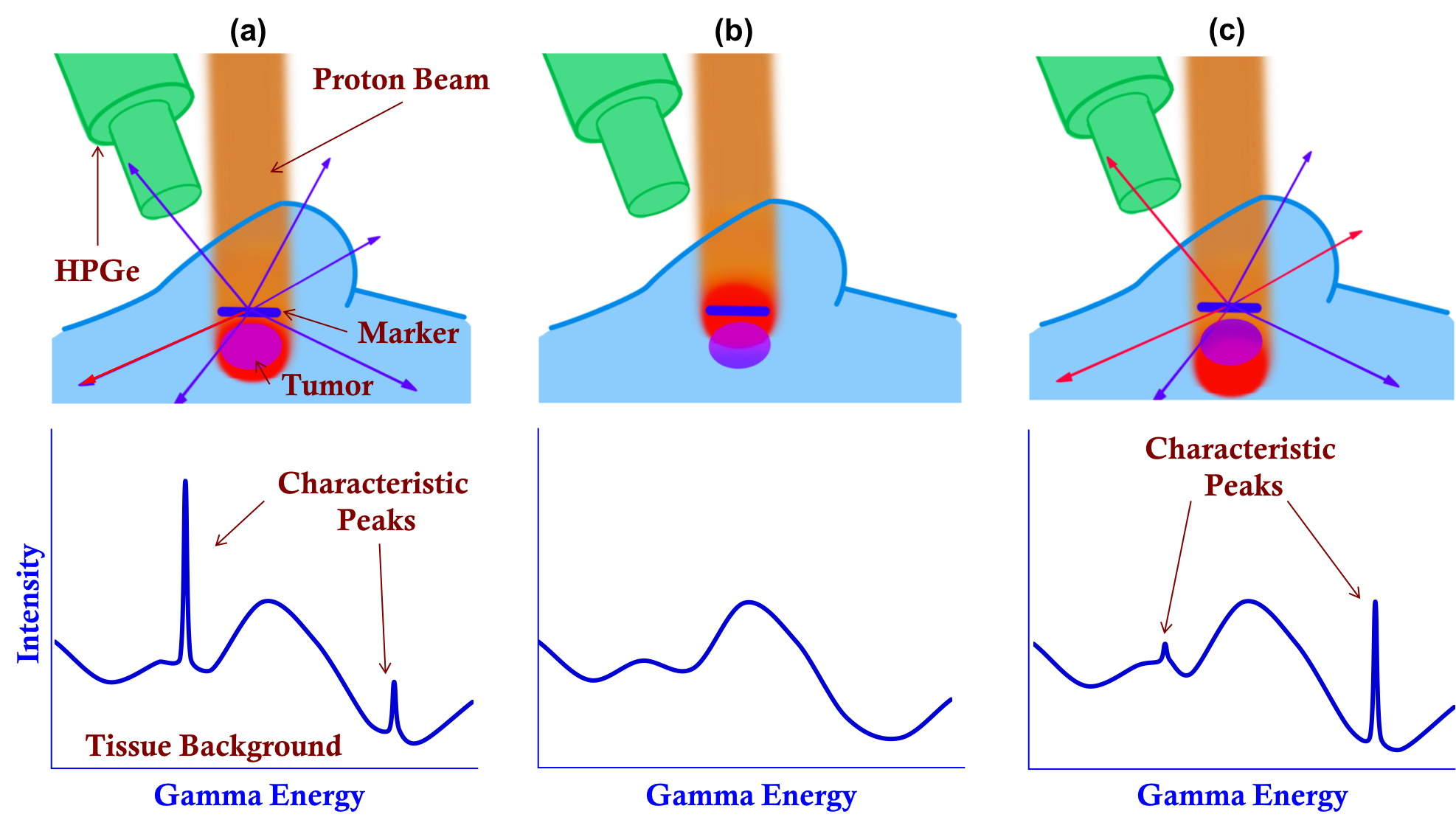}
  \caption{Illustration of proton dose profile and corresponding $\gamma$ spectrum achieved in (a) an ideal treatment scenario, (b) a treatment scenario in which the beam falls short of the expected range, such that no $\gamma$-ray line is observed as the proton energy incident on the HTM is below the threshold for fusion-evaporation reactions, and (c) a treatment scenario in which the the beam stops beyond the expected range, such that a different ratio of the two characteristic $\gamma$ lines is observed.}\label{fig:concept}
\end{figure}

Figure \ref{fig:concept} illustrates this concept in three different treatment scenarios. The main advantages of this method are the energy-dependent cross sections of the fusion-evaporation channels, and the specificity of $\gamma$ rays produced through these interactions to the marker isotope. 

\begin{figure}[!ht]
  \centering
  \subfloat[]{\includegraphics[width=0.3\textwidth]{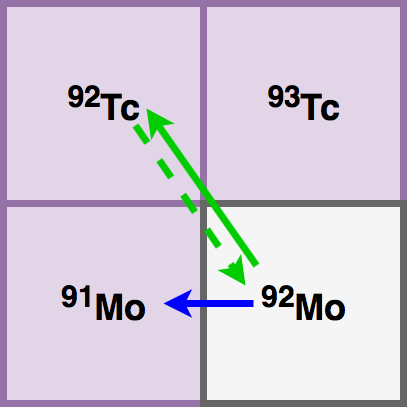}\label{fig:92mo}}
  \hfill
  \subfloat[]{\includegraphics[width=0.37\textwidth]{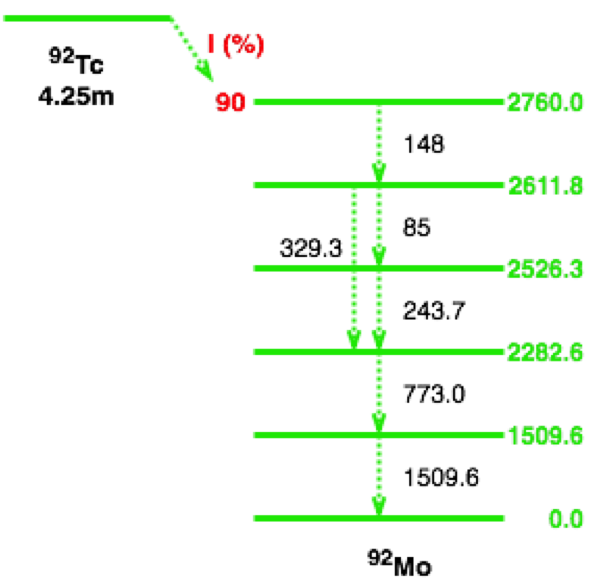}\label{fig:92tc}}
    \hfill
  \subfloat[]{\includegraphics[width=0.30\textwidth]{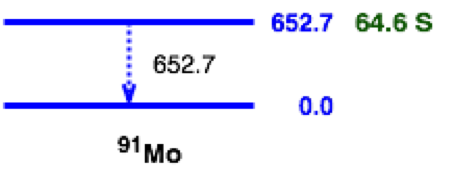}\label{fig:91mo}}
  \caption{(a) Key low-energy reaction pathways from $^{92}$Mo proton activation. In green, the (p,n) path into $^{92}$Tc is shown with the solid arrow, and the subsequent $\beta$-decay into $^{92}$Mo is indicated with the dotted arrow. In blue, the (p,pn) path into $^{91}$Mo is shown. A partial level scheme for the decay of $^{92}$Tc is shown in (b), indicating the energies of relevant $\gamma$ rays (black) and levels (green) in keV. The (p,pn) reaction populates an isomeric state of $^{91}$Mo, whose partial level scheme is similarly depicted in (c). }\label{fig:reactions}
\end{figure}

At proton energies near the Bragg Peak, $^{92}$Mo most commonly undergoes one of two reactions, as illustrated in Figure \ref{fig:reactions}. 

The first, $^{92}$Mo(p,n)$^{92}$Tc, results in the emission of a 773.0(3) keV $\gamma$ ray through the $\beta$-decay of $^{92}$Tc with a half life of 4.25(15) minutes \cite{Baglinpn}. The second reaction of interest is $^{92}$Mo(p,pn)$^{91\text{m}}$Mo. $^{91}$Mo has a low-lying isomeric state that is populated through this reaction, and results in the emission of a 652.9(1) keV $\gamma$ ray with a half life of 64.6(6) seconds \cite{Baglinpd}. The characteristic $\gamma$ rays from these two reactions are suitable indicators of marker activation because they are emitted with high intensity, and the decays responsible for their emission have half lives on the scale of typical fraction delivery times.

The clinical feasibility of HTM RV strongly depends on the half lives of these $\gamma$ rays. Due to the high intensity of prompt $\gamma$ rays present during beam delivery, the ability to measure the decay of the HTM reaction products after the beam has been switched off is a powerful background-suppression tool. By periodic toggling the beam during treatment in order to measure the signal from the marker once the prompt $\gamma$ background has subsided, a greatly amplified SNR can be achieved. In addition, one advantage of the much lower $\gamma$ rates of offline spectroscopy is the possibility of using high purity Germanium (HPGe) detectors, which offer superior energy resolution on the order of 0.2\% at 1.173 MeV \cite{Szymanska2008}. 

Although using $\gamma$ spectroscopy for RV has been explored before, the concept of looking at delayed characteristic $\gamma$ rays from fusion-evaporation reactions in PT is a new method that has not yet been investigated to our knowledge. Unlike PET and PG RV, the beam energy measured using HTM RV is model-independent as it does not rely on comparison to MC simulations.

\section{Methods}

\begin{figure}[!ht]
  \centering
  \includegraphics[width=0.4\textwidth]{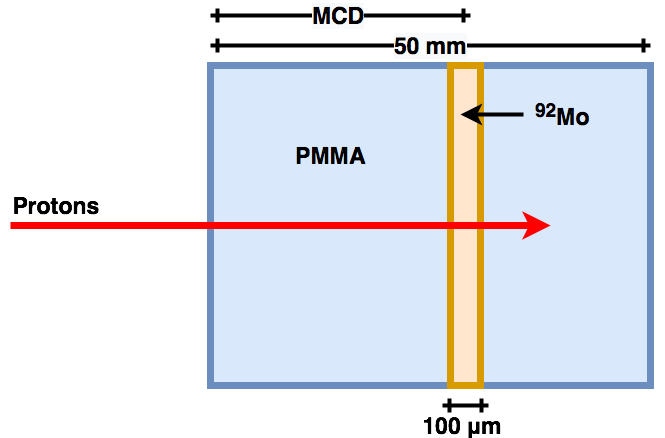}
  \caption{Sketch of 2D cross section of 3D simulation geometry. The tissue-equivalent box is indicated in blue, and the marker region is indicated in orange. The primary protons are generated outside the box, in vacuum. Here, the value MCD represents the marker centre depth, which is varied in order to simulate different treatment scenarios. }\label{fig:setup}
\end{figure}

In order to determine the feasibility of HTM RV, we make use of GEANT4 (v.~10.02), a C$^{++}$-based platform for simulating the passage of particles through matter \cite{Agostinelli2002}. The simulation geometry used in this work is illustrated in Figure \ref{fig:setup}. 

The target geometry consists of an outer box, containing a \SI{100}{\micro\meter}-thick marker region. The box is composed of Poly(methyl methacrylate) (PMMA), which is included in the GEANT4 material database as ``G4\_PLEXIGLASS''. PMMA is often used in radiological phantoms because of its similar density, composition, and stopping power to human soft tissue \cite{Lourenco2017}. Its purpose in this simulation geometry is to provide a $\gamma$ background similar to what would be expected if the marker was located inside a the soft tissue of a phantom or a patient. The marker region is composed of a custom isotope having the nuclear properties of $^{92}$Mo (Z = 42, A = 92). A material density of 10.28 gcm$^{-3}$ was used, to approximate the density of naturally-abundant molybdenum. The depth of the marker within the box was selected to be comparable to that of a typical breast tumour \cite{wang2014}. 76 MeV protons were generated in vacuum (``G4\_Galactic'') outside the target as primary events, and given momentum in the direction of the target geometry. This energy was selected such that the protons would be fully stopped inside the target, between the marker region and the far edge of the PMMA box.

 A modular physics list was created, using QGSP\_BIC as a base due to its applicability in the study of primary protons whose energy is below 10 GeV. In addition, the G4RadioactiveDecayPhysics constructor was added in order to accurately simulate the decay of long-lived reaction products, and the G4HadronPhysicsQGSP\_BIC\_AllHP constructor was included for the implementation of TENDL cross sections in the simulation. The TENDL 1.3.2 library for proton-capture cross sections \cite{KONING2012} was used as it corresponded more closely to experimental results than the default cross sections \cite{Burbadge2019}. In addition, the emission rate of the characteristic $\gamma$ ray of interest from $^{91\text{m}}$Mo was added artificially. When a $^{91}$Mo nucleus is detected as a reaction product, there is a 10\% probability that a $\gamma$ ray entry is created with an energy of 652.9(1) keV and an exponentially distributed randomized time stamp with a half life of 64.6 seconds. The probability of such a $\gamma$ ray being created was set to 10\% in order to reproduce the relative population of 653 keV $\gamma$ rays to 773 keV $\gamma$ in a similar experimental setup \cite{Burbadge2019}. This value is consistent with the results of the aforementioned TALYS calculation, which shows that, at 22.75 MeV, 20.7\% of the total (p,pn) reactions result in the production of the isomeric state, $^{91\text{m}}$Mo. Combining this information with the $\gamma$ intensities found on NNDC \cite{Baglinpd}, we calculate a maximum emission rate of 10.01(21)\% of the 652.9(7) keV $\gamma$ ray in a (p,pn) reaction. Similarly, the average relative emission rate between 20 MeV and 30 MeV was calculated to be 9.1(13)\%.    

To simplify the simulations, a realistic beam energy spread and spatial distribution, as well as detector efficiencies and resolution, were not included. Instead, all $\gamma$ rays are recorded upon exiting the box, and detector resolution and efficiency are applied during the data analysis using ROOT 6.10.8 \cite{root}.
The GEANT4 simulation tracked the instantaneous passage of $10^8$ primary protons through the target geometry, recording the energy and timing of all $\gamma$ rays exiting the box. Based on an intrinsic detector efficiency of 10\%, and a geometric efficiency of 33\%, the number of primaries simulated produce statistics that are equivalent to those of a 2.7 Gy fraction dose delivered with a pencil beam. In order to more accurately portray beam intensities, a basic beam timing microstructure was implemented post-simulation in ROOT to approximate a treatment beam current. In order to simulate pulsed-beam delivery, a beam timing macrostructure was also implemented in ROOT, allowing for the periodicity of the beam pulses to be selected post-simulation.

\section{Simulation Results}
\begin{figure}[!ht]
  \centering
  \includegraphics[width=0.5\textwidth]{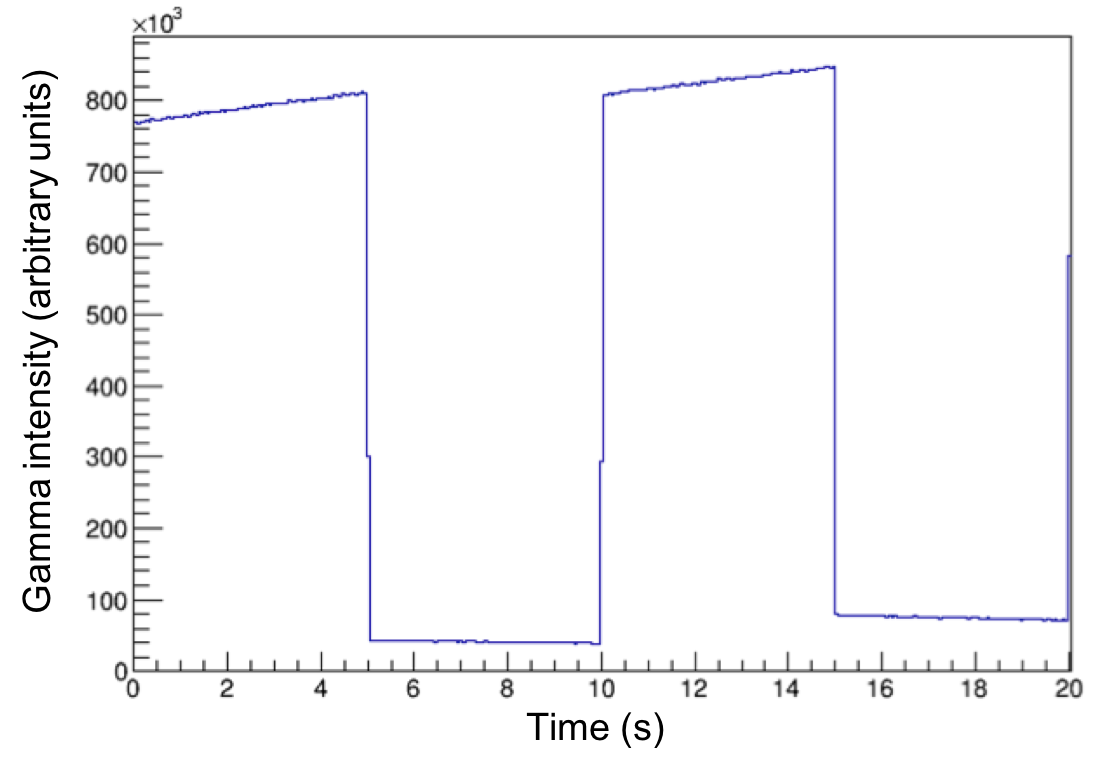}
  \caption{Simulation results of $\gamma$ intensity as a function of time for periodic beam switching at 5 second intervals. This illustrates the immediacy and magnitude of the reduction in background rates.} \label{fig:beam-timing}
\end{figure}

Figure \ref{fig:beam-timing} shows the simulated $\gamma$-ray intensity pattern observed using a periodic beam macrostructure of 5 seconds on-beam, followed by 5 seconds off-beam, at a beam current of 0.75 nA  with $10^8$ events.

\begin{figure}[!ht]
  \centering
  \includegraphics[width=0.7\textwidth]{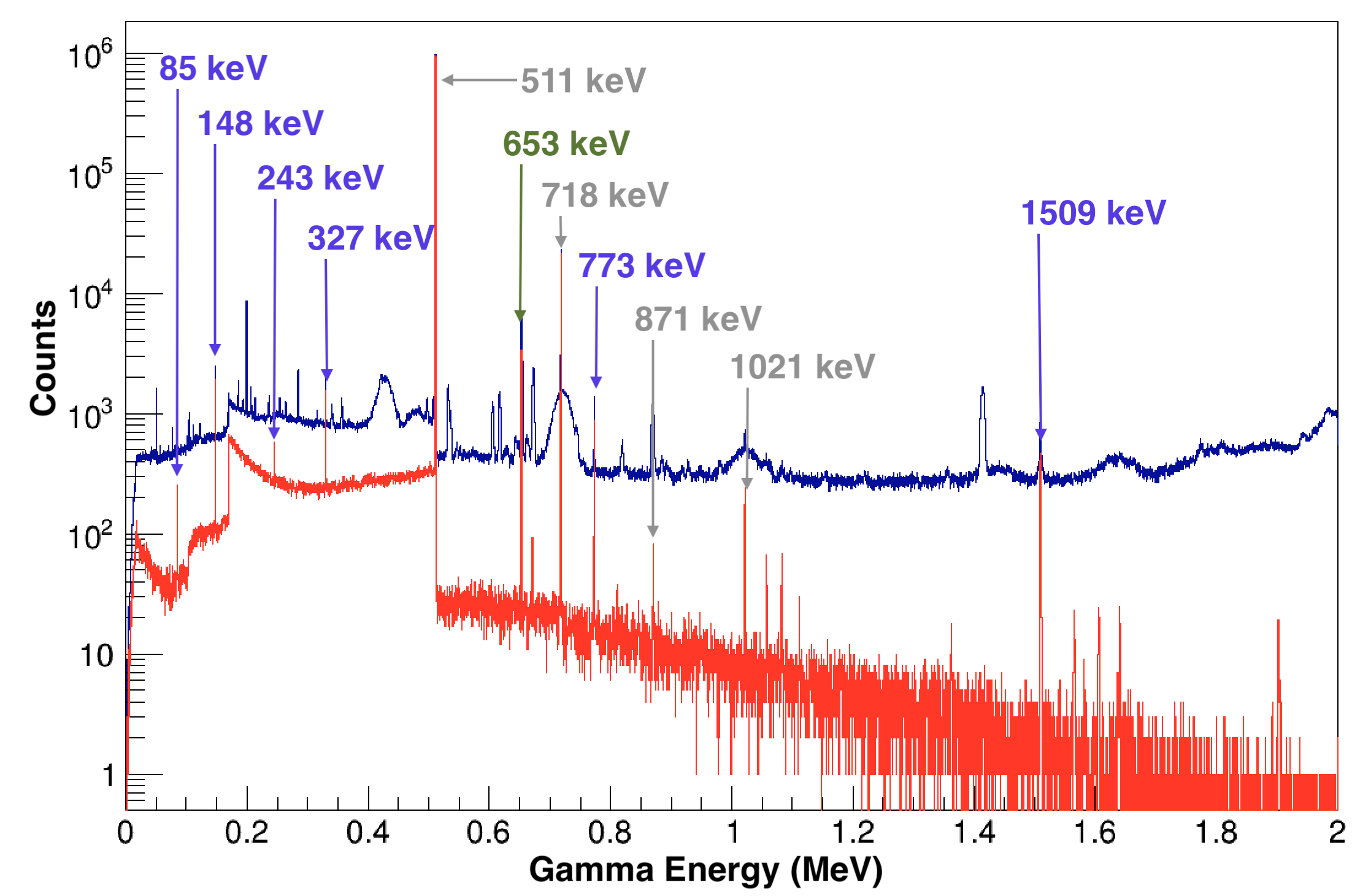}
  \caption{Simulated $\gamma$ spectrum for $10^8$ protons impinging on $^{92}$Mo foil embedded in PMMA. The blue line shows a histogram of all $\gamma$ rays produced, while the red line shows a histogram of only the $\gamma$ rays produced in the off-beam window. The peaks labelled in violet correspond to the characteristic $\gamma$ rays produced as a result of the decay of $^{92}$Tc, which has a half life of 4.25 minutes, and the peak labelled in green is produced through the $\gamma$ decay of $^{91\text{m}}$Mo. The peaks labelled in grey correspond to prominent background peaks from the tissue-equivalent plastic (PMMA). A Gaussian smear replicating an energy resolution of 0.2\% was applied to these spectra. }\label{fig:fullbkgd-all}
\end{figure}

\begin{figure}[!ht]
  \centering
  \includegraphics[width=0.7\textwidth]{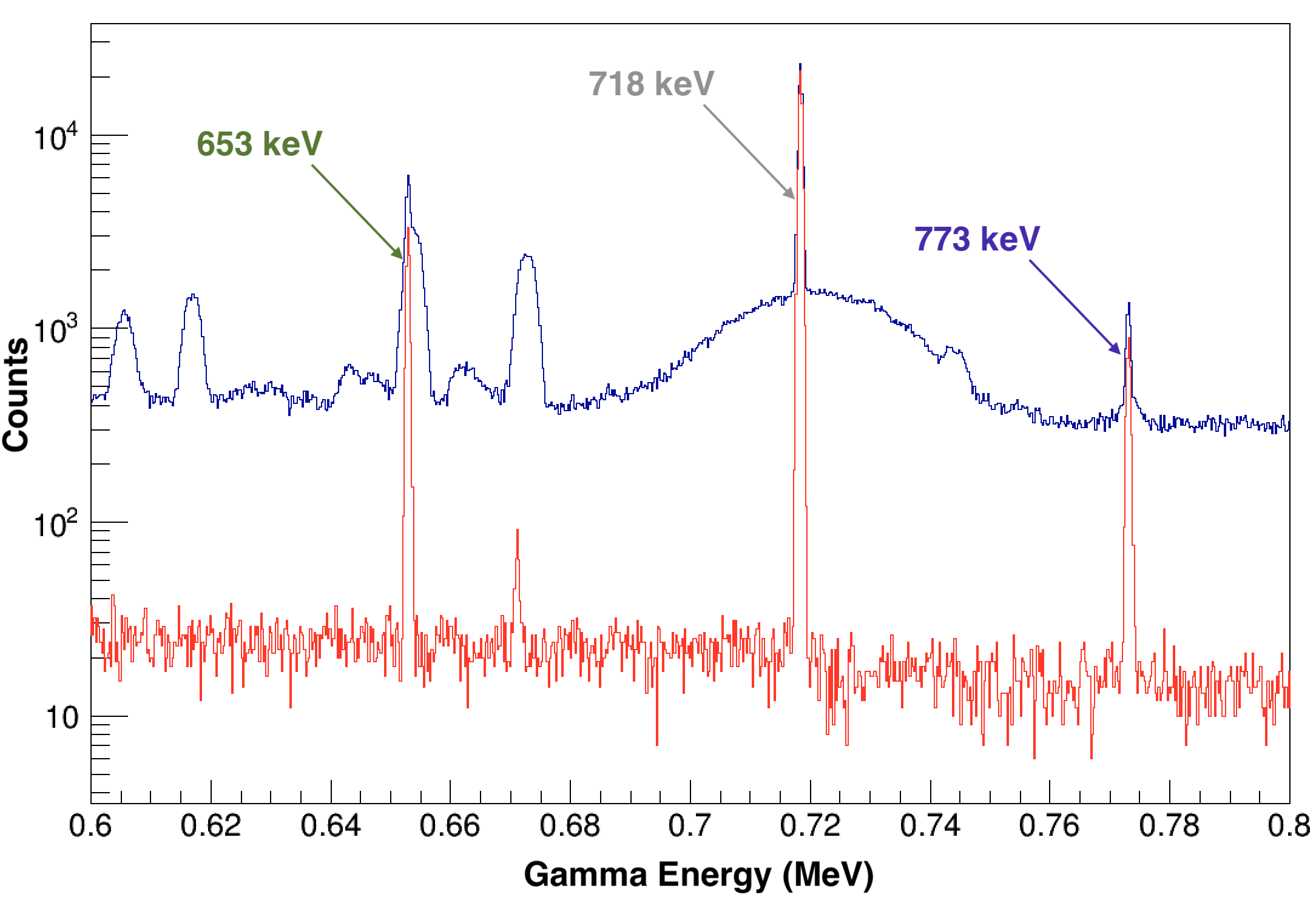}
  \caption{Same plot as depicted in Figure \ref{fig:fullbkgd-all}, focused on the region containing peaks of interest. }\label{fig:fullbkgd-zoom}
\end{figure}

The majority of the $\gamma$-ray and neutron background is produced on-beam. By periodically stopping the beam and performing $\gamma$-ray spectroscopy off-beam, the SNR of the characteristic peaks can be drastically improved compared to on-beam measurements. Due to the longer rise and decay times of their pulse signals, HPGe detectors are more susceptible to large dead times at high $\gamma$-ray rates (above 100kHz) \cite{ortec2003}. Thus, the reduction in $\gamma$-ray count rates achieved by measuring off-beam is a crucial aspect of this RV method. The impact of the background reduction in the off-beam window on the resulting simulated $\gamma$ spectrum can be seen in Figures \ref{fig:fullbkgd-all} and \ref{fig:fullbkgd-zoom}. Figure \ref{fig:fullbkgd-zoom} shows the energy region containing the two peaks of interest for this method. The 773 keV $\gamma$ ray from the decay of $^{92}$Tc, and the 653 keV $\gamma$ ray from the decay of $^{91\text{m}}$Mo. These gamma rays were selected as they have large branching ratios, and are thus emitted at high intensities from the HTM. In addition, the similar energies of these two gamma rays reduces the impact of energy-dependent detector efficiency and tissue absorption on the measurement accuracy.


Here, a comparison of the simulated $\gamma$-ray spectra obtained by measuring both prompt and delayed $\gamma$ rays, as opposed to measuring only in the off-beam time window is shown. The simulated data replicates the timing and energy resolution of a HPGe detector. The result is a dramatic decrease in tissue background, and a significant increase in the SNR for the $\gamma$ rays of interest. 

\begin{figure}[!ht]
  \centering
  \subfloat[]{\includegraphics[width=0.45\textwidth]{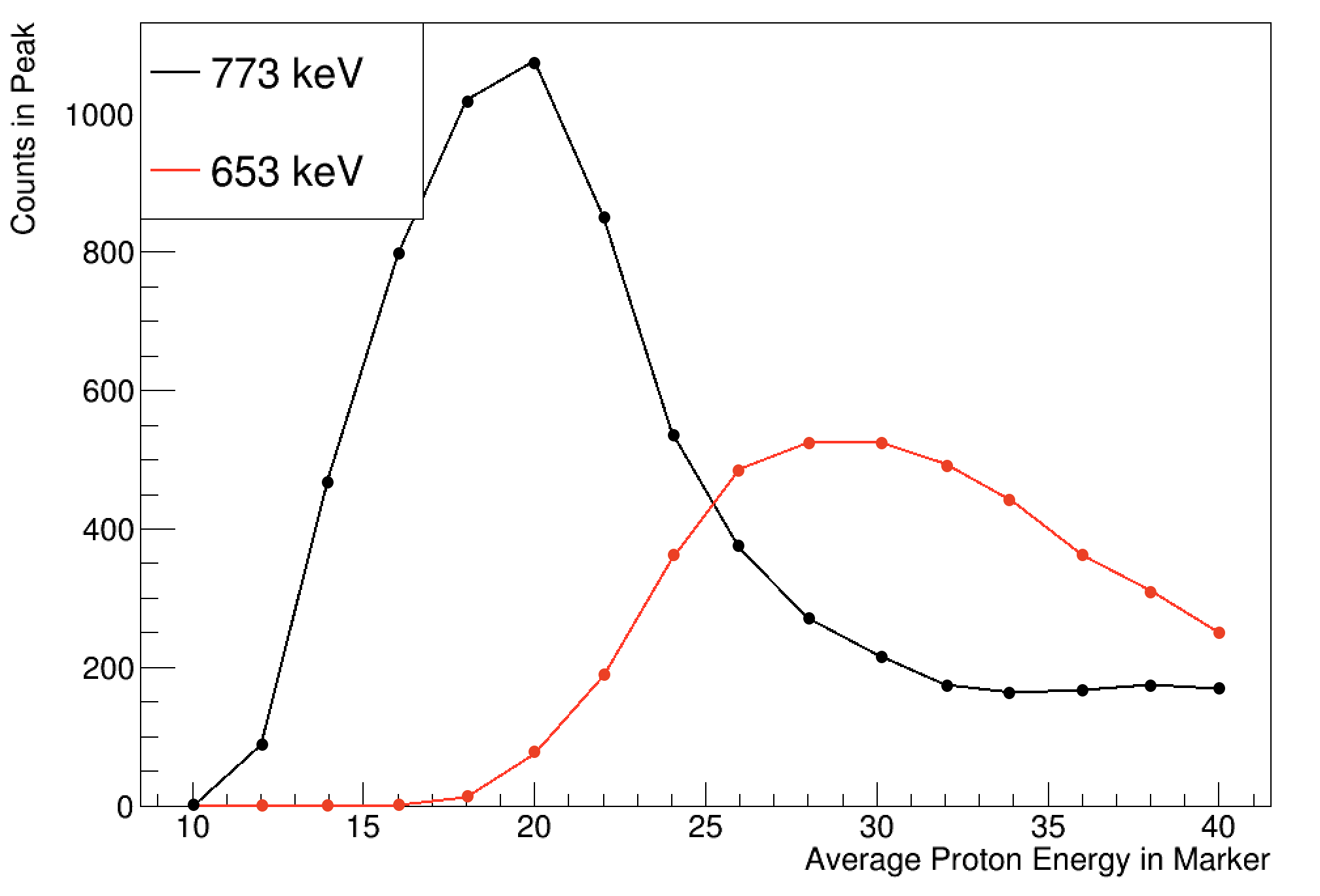}\label{fig:talys}}
  \hfill
  \subfloat[]{\includegraphics[width=0.45\textwidth]{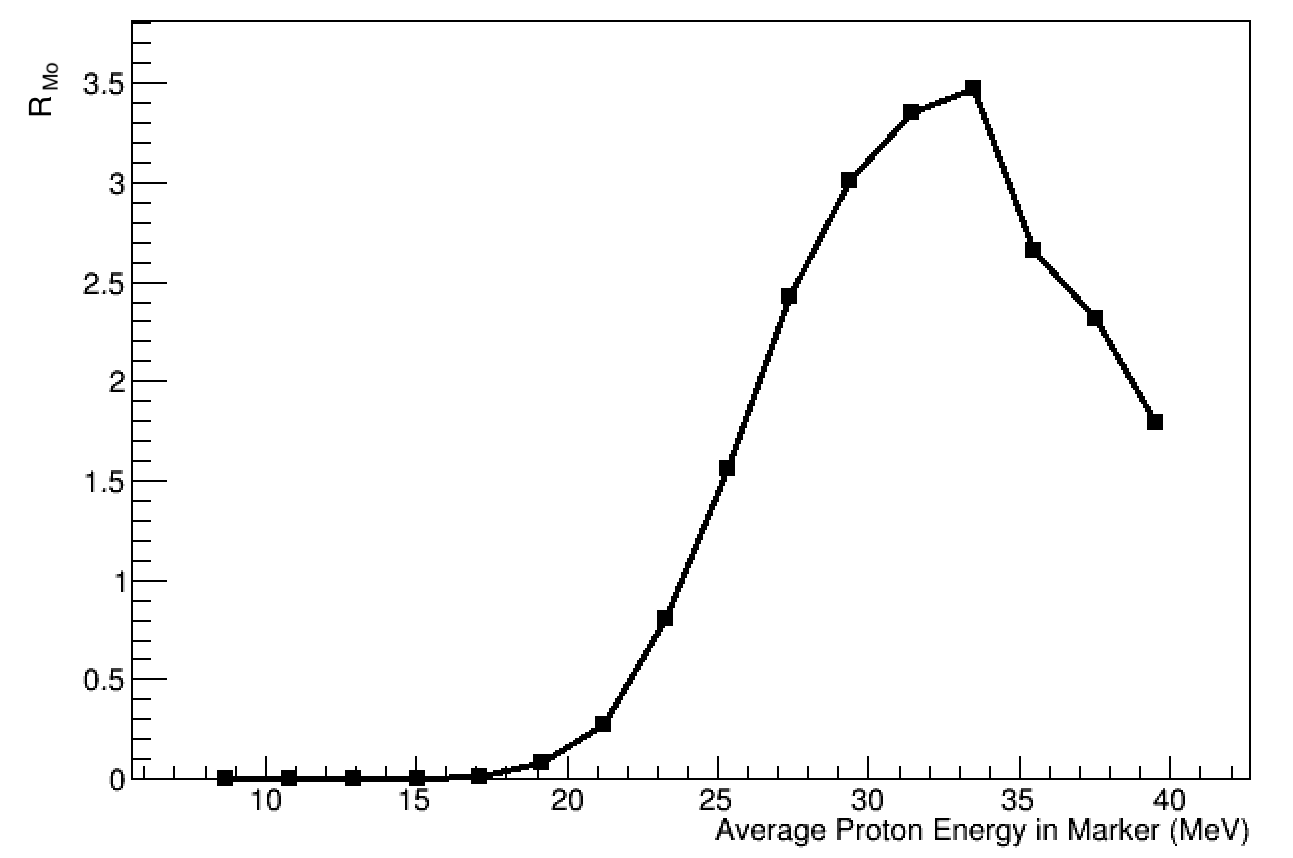}\label{fig:ratio}}
  \caption{ (a) Simulated counts of $\gamma$ rays of interest (773.0(3) keV and 652.9(1) keV) as a function of average proton energy in the marker. (b) Same data, with y-axis expressed as a ratio of the intensity of the 652.9(1) keV peak to that of the 773.0(3) keV peak, $R_{\text{Mo}}$. Note that as this data is directly extracted from the simulation, with no energy resolution applied, error bars have not been included.}\label{fig:ratio-sim}
\end{figure}

Primary protons of various energies were generated and impinged on a \SI{100}{\micro\meter}-thick foil of $^{92}$Mo in vacuum (G4\_Galactic). In this low-background simulation environment, the intensity of the $\gamma$ rays of interest emitted from the marker were extracted with high precision, with no gaussian energy smear, to produce a calibration curve. The $\gamma$-ray intensities for the two peaks of interest were plotted in Figure \ref{fig:talys} as a function of the mean proton energy inside the marker region. Figure \ref{fig:ratio} depicts the same simulation data, but the $\gamma$-ray intensity is expressed as a ratio of the peak integrals of the 652.9(1) keV peak to the 773.0(3) keV peak, $R_{\text{Mo}}$. Expressing the $\gamma$-ray intensity as $R_{\text{Mo}}$ makes the curve shown in figure \ref{fig:ratio-sim} independent of dose, beam intensity, and deadtime. The simulation results depicted in Figure \ref{fig:ratio-sim} were obtained without any Gaussian smear applied to the time and energy of the $\gamma$ rays, nor any energy-dependent detector efficiency considerations. However, since the $\gamma$ rays of interest are very close in energy, the energy dependence of the energy resolution has a minimal impact on calculations.

In order to relate $R_{\text{Mo}}$ to the range of the beam, the relationship between a proton's energy and its remaining range in tissue must be established. For this purpose, we make use of a numerical solution to the velocity profile of protons developed by \citeasnoun{Martinez2019}.



\begin{figure}[!ht]
  \centering
  \subfloat[]{
  \label{fig:zve}
  \includegraphics[width=0.45\textwidth]{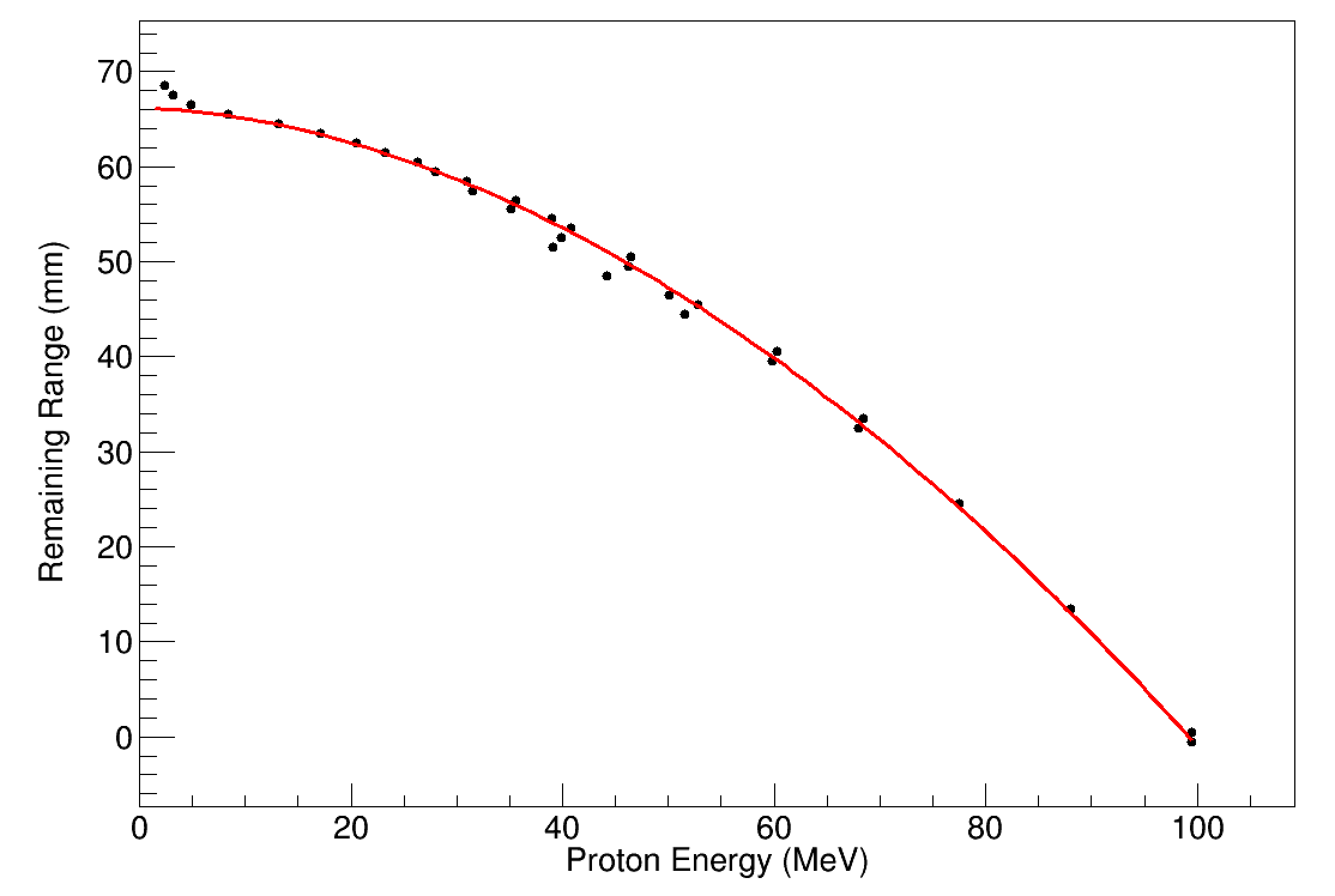} 
  }
  \hfill
  \subfloat[]{
   \label{fig:ratio-range}
  \includegraphics[width=0.45\textwidth]{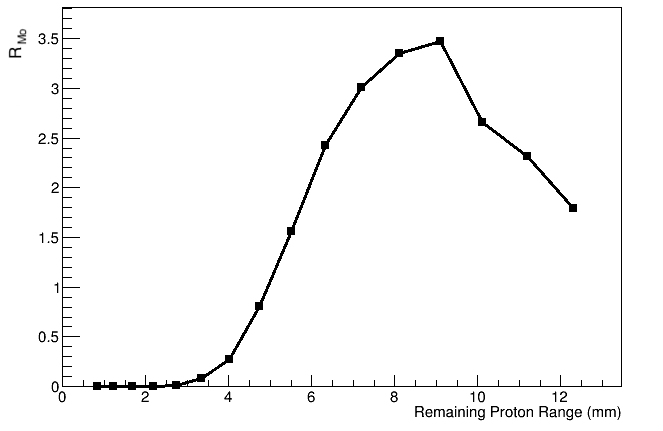}
  }
  
  \label{fig:range-conversion}
\end{figure}

The numerical solution was applied to simulation data depicting depth of a proton in PMMA as a function of its energy at that depth, yielding the plot shown in Figure \ref{fig:zve}. 

Simulations of a 76 MeV proton beam impinging on an HTM embedded in PMMA were performed at various marker depths within the plastic in order to characterize the range sensitivity of HTM RV. In all simulations, $10^8$ proton events were generated and the reaction products were allowed to decay for 600 seconds. The simulated ratio, $R_{\text{Mo}}$, was compared to Figure \ref{fig:ratio-range} in order to extract the remaining range of the proton beam in the PMMA. It should be noted that due to the limited range of energies for which a non-zero ratio can be measured, the projection of $R_{\text{Mo}}$ alone onto the range axis does not provide a unique solution. However, if the HTM is positioned in the most sensitive region, approximately \SI{6}{\milli\meter} from the distal edge of the beam, then a typical treatment precision \cite{Paganetti2015} should be sufficient to avoid any ambiguity in the extracted range for a given $R_{\text{Mo}}$. 

\begin{figure}[!ht]
  \centering
  \includegraphics[width=0.6\textwidth]{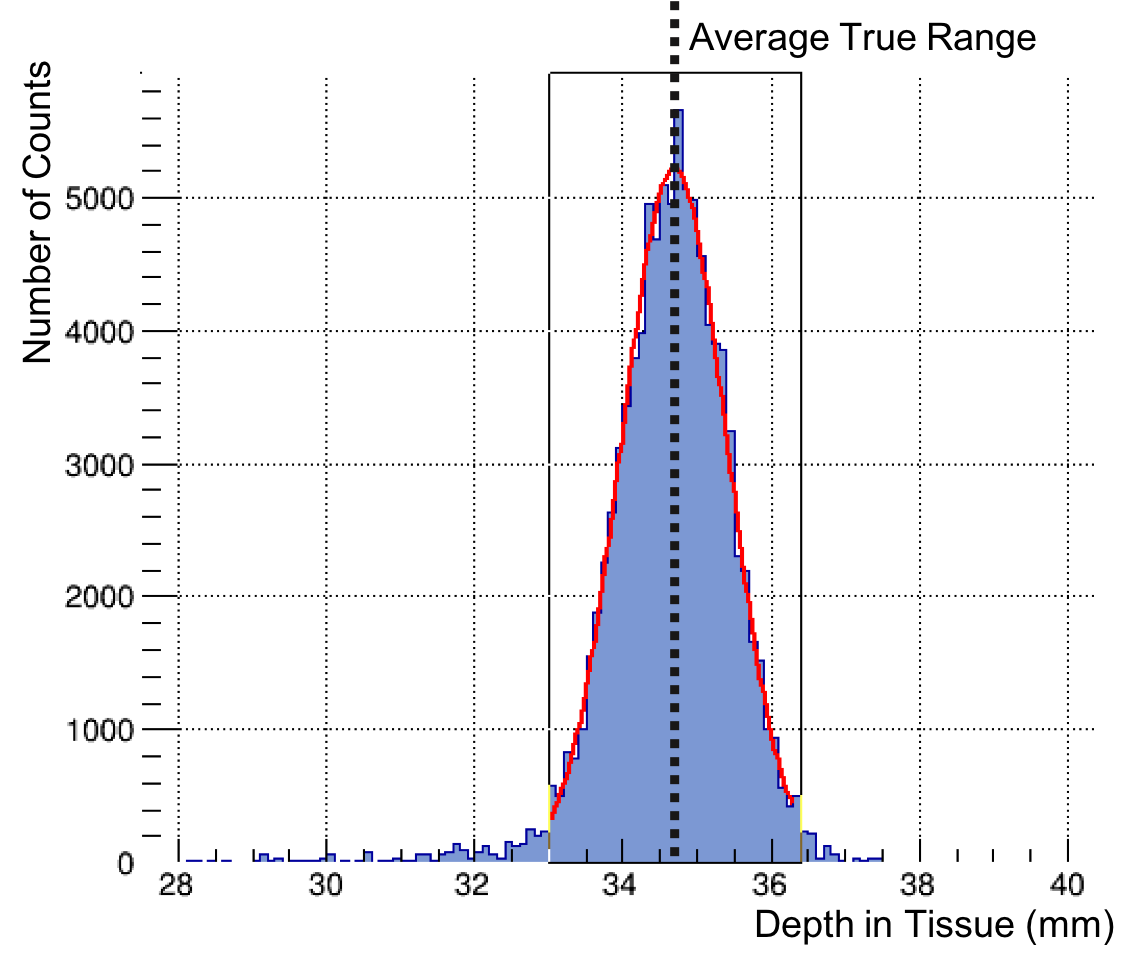}
  \caption{Example histogram of depth in PMMA at which 76 MeV protons come to rest. A Gaussian fit was used in order to extract the mean of this distribution.}\label{fig:proton-range}
\end{figure}

The ``true remaining range'' of the proton beam was determined by plotting the average depth after the HTM centre at which simulated protons come to rest, then extracting the mean by fitting the resulting Gaussian distribution, as shown in Figure \ref{fig:proton-range}.

\begin{figure}[!ht]
  \centering
  \includegraphics[width=0.6\textwidth]{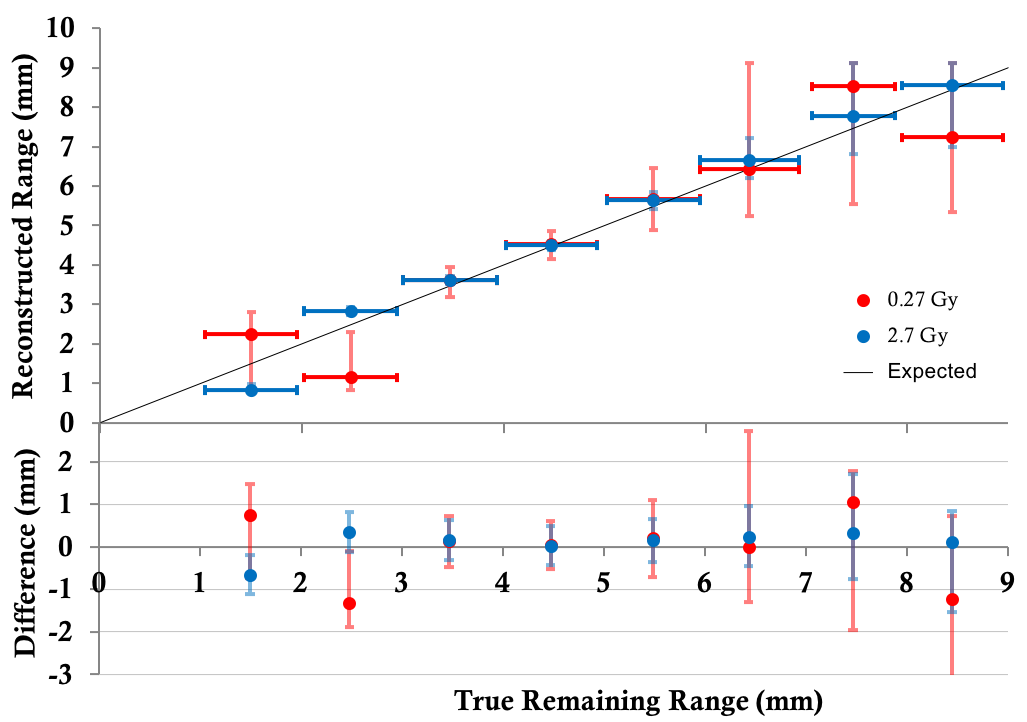}
  \caption{(top) Beam range calculated using simulated $R_{\text{Mo}}$, compared to true range of protons after the marker, extracted from simulation data. The straight line indicates the expected value for these data points, i.e. the ideal case in which the true range and extracted range are identical. (bottom) Difference between calculated and expected values for proton range. The standard deviations of these data sets from expectations are $\sigma_{2.7 Gy} =$ \SI{0.32}{\milli\meter}, and $\sigma_{0.27 Gy} =$ \SI{0.84}{\milli\meter}.}\label{fig:range sensitivity}
\end{figure}

The extracted remaining range from Figure \ref{fig:ratio-range} was compared to the true remaining range in Figure \ref{fig:range sensitivity}.  The data plotted in red show the same results for delivery of only 0.27 Gy instead of 2.7 Gy, with the aforementioned detector efficiencies. Using a thicker marker would result improved peak SNR at the cost of a reduced precision in range due to the larger spread of proton energies interacting with the marker.

\section{Discussion}

The results of this feasibility study have shown that the intensity ratio of the competing reactions of interest, $R_{\text{Mo}}$, is a highly sensitive indicator of proton energy inside the marker. The $R_{\text{Mo}}$ extracted from this method can be directly translated to the beam’s range relative to the marker as a form of in-vivo RV in real time, without the need to refer to MC simulations. The simulation results obtained here suggest that an SNR of up to 38(6) can be achieved for the 773.0(3) keV peak in the delayed-$\gamma$ spectrum with an energy resolution typical for HPGe detectors. Due to the photon-energy dependence of the linear absorption coefficient of tissue, the selection of $\gamma$ rays with similar energies is favourable for this analysis in order to minimize discrepancies. For a scenario in which the tumour is located deep inside the tissue, literature values \cite{salehi2015,Biswas2016} suggest that for the same number of emitted 600 keV and 800 keV $\gamma$ rays passing through 20(1) cm of soft tissue, an uncertainty of 0.85\% can be expected in $R_{\text{Mo}}$. This uncertainty is dependent to the depth of the marker inside the tissue and would thus be much smaller for tumours located closer to the surface of the skin.

The simulation results expressed in Figure \ref{fig:range sensitivity} indicate that HTM RV provides a precise reconstruction of the range of the beam with no offset. As indicated by the two plotted data sets, the range in which this method will offer accurate range verification will be dependent on the SNR that can be achieved clinically. For the setup investigated in this work, the placement of the marker within the tissue that will offer the most sensitive response is approximately 4.5 mm in front of the point at which the beam should be stopped. At this position, and with the statistics used to generate the data points representing the full fraction dose of 2.7 Gy, the marker's response to proton activation will allow for direct measurement of the difference in the beam's range from its target range, up to a difference of \SI{2}{\milli\meter} in either direction. Outside this range, the reduced cross section in one of the two peaks of interest results in larger uncertainties in the reconstructed range. Since typical clinical precision of the range is generally under \SI{2}{\milli\meter}, the application of HTM RV in the region of highest sensitivity can provide a unique measurement of the beam's range, with no degeneracy within the range of beam energies expected. In addition, the reduction in beam energy induced by the passage of the proton beam through a \SI{100}{\micro\meter} HTM is minimal and can be well characterized to reduce impact on fraction delivery.

The beam pulsing technique utilized in obtaining this spectrum allows for a signal enhancement of over an order of magnitude at the cost of an extension of the fraction delivery duration. However, in specific cases such as the radiological treatment of breast cancers, the timing of the fraction delivery may already be limited by the patient’s breathing cycle, allowing for the implementation of HTM RV with minimal impact on delivery times. It should be noted that this simulation does not take into consideration the $\gamma$ and neutron background in the treatment room, nor the effects of Compton scattering.

In general, PT fractions are delivered with a range of beam energies to ensure full coverage of the CTV. Our method is well-suited for pencil-beam PT, where the beam energy is well defined and can be varied over the course of the fraction delivery. The results of the 0.27 Gy simulation suggest that our method could be used prior to fraction delivery as a method of quality control, as well as during treatment for online monitoring. The HTM would be surgically implanted near the CTV before treatment \cite{Tran2012}. The position of the implanted HTM inside the tissue can be precisely measured prior to treatment with a CT scan. Using a small portion of the dose delivered with a mono-energetic beam, the RV method outlined in this work can provide a high-precision measurement of the proton energy in the HTM before delivering the remainder of the fraction. Thereafter, the signal from the HTM will continue to provide real-time range feedback for the remainder of the fraction delivery. This particular method also has applications in RV for FLASH PT, a promising new treatment method in which the entire fraction dose is delivered at rates exceeding 40 Gy/s \cite{vandewater2019}. It has been shown that FLASH dose delivery results in increased therapeutic index.

\section{Conclusion}

The in-vivo RV method outlined in this work is able to determine the absolute remaining range of the beam inside the patient more precisely than any other RV method currently in clinical use, without the need for comparison to complex MC simulations, and is largely independent of tissue composition. The energy of the beam in the HTM is directly measured through the peak intensity ratio, $R_{\text{Mo}}$, and the remaining sub-cm range of the beam is calculated with a very small uncertainty. With a simple and relatively cost-effective measurement setup consisting of a small $\gamma$-ray detector array and a digital data acquisition system, this method is highly accessible for clinical use. With acquisition times much shorter than PET, and range measurements taking place prior to the completion of fraction delivery, this method is able to provide feedback on treatment accuracy in real time, on a sub-mm scale. Future work will include the investigation of the biological impact of the implanted markers, different HTM candidates, expansion of the simulation to include more realistic background levels, as well as optimization of detector geometries. 

\section*{Acknowledgements}
We acknowledge the support of the CIHR, NSERC and SSHRC (under Award No. NFRFE-2018-00691). TRIUMF receives federal funding via a contribution agreement with the National Research Council of Canada.

\section*{Bibliography}

\bibliographystyle{dcu}
\bibliography{sample}

\end{document}